\documentclass[prb,floatfix,twocolumn,showpacs,amsmath,amssymb]{revtex4}
\usepackage{graphicx}
\usepackage{epstopdf}
\usepackage{epsfig}
\usepackage{dcolumn}
\usepackage{bm}

\begin{document}

\title{Direct evidence for a gapless $Z_2$ spin liquid by frustrating N\'eel 
antiferromagnetism}
\author{Wen-Jun Hu,$^1$ Federico Becca,$^1$ Alberto Parola,$^2$ and 
Sandro Sorella$^1$}
\affiliation{
$^1$ Democritos Simulation Center CNR-IOM Istituto Officina dei Materiali and 
International School for Advanced Studies (SISSA), Via Bonomea 265, 34136 
Trieste, Italy \\
$^2$ Dipartimento di Scienza e Alta Tecnologia, Universit\`a dell'Insubria, 
Via Valleggio 11, I-22100 Como, Italy}

\date{\today}

\begin{abstract}
By direct calculations of the spin gap in the frustrated Heisenberg model on 
the square lattice, with nearest- ($J_1$) and next-nearest-neighbor ($J_2$) 
super-exchange couplings, we provide a solid evidence that the spin-liquid
phase in the frustrated regime $0.45 \lesssim J_2/J_1 \lesssim 0.6$ is gapless.
Our numerical method is based on a variational wave function that is 
{\it systematically} improved by the application of few Lanczos steps and 
allows us to obtain reliable extrapolations in the thermodynamic limit. 
The peculiar nature of the non-magnetic state is unveiled by the existence of 
$S=1$ gapless excitations at $k=(\pi,0)$ and $(0,\pi)$. The magnetic transition
can be described and interpreted by a variational state that is built from 
Abrikosov fermions having a $Z_2$ gauge structure and four Dirac points in 
the spinon spectrum. 
\end{abstract}

\pacs{}

\maketitle

{\it Introduction --}
During the ``Valence-Bond-Solid era'' most of the community working on 
highly-frustrated magnets believed that quantum spin liquids could not exist 
as true ground states of microscopic models and some kind of valence-bond
order would have taken place in non-magnetic insulators (thus leading to 
trivial band insulators). Now, we are presently living the more exciting 
``Quantum-Spin-Liquid era'', where a plethora of different spin-liquid states 
are proposed as ground states of various magnetic systems, both theoretically 
and experimentally.~\cite{balents2010} The turning point was marked by the 
discovery that stable gapped spin liquids may be found in effective 
low-energy Hamiltonians, which are based upon the so-called quantum dimer 
models~\cite{moessner2001} or strong-coupling expansions.~\cite{balents2002} 
Since then, three main directions can be identified to study quantum spin 
liquids. The first one is the definition of {\it ad hoc} Hamiltonians that can
be exactly solved to have a cartoon picture of the exotic properties 
expected in generic systems (e.g., topological degeneracy and fractional 
excitations).~\cite{kitaev2003,kitaev2005} The second one is the classification
of different spin-liquid states according to hidden symmetries (i.e., beyond 
the Ginzburg-Landau description); examples may be given by the 
projective-symmetry group,~\cite{wen2002} tensor states,~\cite{chen2011a}
or cohomology.~\cite{chen2011b,hermele2012,ran2012} Finally, the third and more
pragmatic one is to perform numerical simulations on Heisenberg or Hubbard 
models, in order to gain evidence that stable spin-liquid phases may indeed 
exist.~\cite{capriotti2001,yan2011,jiang2012,iqbal2013,meng2010,sorella2012}

In this Letter, we take the latter point of view and investigate the 
$J_1{-}J_2$ spin-half Heisenberg model on the two-dimensional square lattice 
by systematically improving accurate variational wave functions, to obtain a 
reliable estimate of the {\it exact} ground state, along with few relevant 
low-energy states. This procedure allows us to extract the spin gap and show 
that a gapless spin-liquid phase exists in the highly frustrated regime.

The Heisenberg $J_1{-}J_2$ model is defined by:
\begin{equation}
{\cal H} = J_1 \sum_{\langle i,j \rangle} {\bf S}_i \cdot {\bf S}_j +
J_2 \sum_{\langle \langle i,j \rangle \rangle} {\bf S}_i \cdot {\bf S}_j,
\end{equation}
where ${\bf S}_i = (S^x_i, S^y_i, S^z_i)$ is the quantum spin operator on the 
site $i$; $\langle \dots \rangle$ and $\langle \langle \dots \rangle \rangle$ 
indicate nearest-neighbor and next-nearest-neighbor sites. Here, we focus on 
the case where both super-exchange couplings are antiferromagnetic and consider 
clusters with $N=L \times L$ sites and periodic boundary conditions.

In the unfrustrated case with $J_2=0$, it is well established that the ground 
state has N\'eel long-range order, with a staggered magnetization that is 
reduced from its classical value, i.e., 
$M \simeq 0.307$.~\cite{sandvik1997,calandra1998} For large values of $J_2$, 
the ground state shows again a collinear magnetic order with pitch vector 
$Q=(\pi,0)$ or $(0,\pi)$. The intermediate regime, around the strongest 
frustration point $J_2/J_1=0.5$, is the most debated one, since the combined
effect of frustration and quantum fluctuations destroys antiferromagnetism and
leads to a non-magnetic ground state. However, the nature of this quantum phase
is still controversial. Since the pioneering 
works,~\cite{doucot1988,read1989,gelfand1989,figueirido1990} it was 
clear that the problem was terribly complicated: many states can be 
constructed with very similar energies but very different physical properties,
e.g., having dimer or plaquette valence-bond order, or being totally
disordered with short- or long-range resonating-valence bond fluctuations.
This is mainly due to the fact that the non-magnetic region of the $J_1{-}J_2$ 
model is relatively small and several generalized susceptibilities may be quite 
large,~\cite{darradi2008} indicating that the ground state is on the verge of 
various instabilities. In this context, there is a convincing evidence that a 
third-nearest-neighbor coupling $J_3$ may drive the system into a valence-bond 
solid.~\cite{mambrini2006}

Recent density-matrix renormalization group (DMRG) results sparked the desire 
of understanding the phase diagram of the $J_1{-}J_2$ model, suggesting
the existence of a true spin-liquid phase.~\cite{jiang2012} In particular, by 
considering cylindrical geometry, results for the singlet and triplet 
gaps provided some evidence for a fully gapped $Z_2$ state in the region 
$0.41 \le J_2/J_1 \le 0.62$, without local broken symmetry.
Moreover, the calculation of the so-called topological entanglement 
entropy $\gamma$ was found to be consistent with the expected value of 
$\gamma=\ln(2)$ for a gapped $Z_2$ spin liquid. The most natural description 
of a fully gapped state is given in terms of the Schwinger boson representation
of the spin operators.~\cite{arovas1988} By performing a full optimization 
of the many-body wave function on small sizes, we showed that this kind of 
bosonic ansatz may qualitatively reproduce some of the DMRG 
results.~\cite{li2012} However, while in the weakly-frustrated regime 
the bosonic ansatz has magnetic order and excellent variational energy, 
for $0.45 \lesssim J_2/J_1 \lesssim 0.6$ a state constructed with Abrikosov
fermions instead of Schwinger boson has better accuracy.~\cite{li2012} In fact,
in a forerunner paper,~\cite{capriotti2001} three of us showed that, within
this kind of fermionic representation, it is possible to have a particularly
accurate description of the ground state in the strongly frustrated regime.
By using the language of the projective-symmetry group (PSG),~\cite{wen2002} 
our variational wave function (dubbed Z2Azz13 in Ref.~\onlinecite{wen2002}) 
has a $Z_2$ gauge structure (implying gapped gauge excitations) but gapless 
spinon excitations with four Dirac points. A gapless spin liquid with 
topological degeneracy has been also suggested by using projected-entangled 
pair states on cylindrical geometry.~\cite{wang2013} These results suggest
that a gapless spin liquid may be competitive with the one proposed by DMRG 
calculations. 

In this Letter, we present numerical calculations based upon a systematic
improvement of the fermionic Z2Azz13 wave function that allow us to extract 
(i) the ground state energy, (ii) the energy of the lowest $S=2$ state, and 
(iii) the energy of a state with $S=1$ and $k=(\pi,0)$ [or $(0,\pi)$], so to 
extract the information about the {\it exact} spin gap. The state 
with $S=1$ and $k=(\pi,0)$ is particularly interesting, since it is certainly 
gapped in the N\'eel phase and it is not expected to play any important role 
in a gapped non-magnetic regime (while it is one of the gapless modes in the 
collinear magnetic phase that appears for large $J_2$ values). On the contrary,
this state is gapless in the Z2Azz13 ansatz (see below for the details). 
One of the main results of this work is to show that this $S=1$ excitation 
becomes indeed gapless in a region around $J_2/J_1=0.5$, and, therefore, 
a spin liquid with gapless triplet excitations both at $k=(\pi,0)$ and 
$(\pi,\pi)$ represents the most natural candidate between the two magnetic
phases characterizing the small and large $J_2/J_1$ regimes.

{\it Numerical method --}
The staring variational wave functions are defined through the mean-field
Hamiltonian for the Abrikosov-fermion representation of the spin 
operators:~\cite{baskaran1987}
\begin{eqnarray}
{\cal H}_{MF} &=& \sum_{i,j,\sigma} t_{i,j} c^\dag_{i,\sigma} c_{j,\sigma} 
+ h.c. \nonumber \\
&+& \sum_{i,j} \eta_{i,j} (c^\dag_{i,\uparrow} c^\dag_{j,\downarrow} + 
c^\dag_{j,\uparrow} c^\dag_{i,\downarrow}) + h.c.,
\label{eq:meanfield}
\end{eqnarray}  
where for each bond $(i,j)$ there are hopping ($t_{i,j}$) and/or pairing 
($\eta_{i,j}$) terms; the mean-field Hamiltonian may also contain on-site 
terms (i.e., chemical potential and/or on-site pairing). Given any eigenstate
$|\Psi_{MF}\rangle$ of the mean-field Hamiltonian, a physical state for the 
spin model can be obtained by a projection of it onto the subspace with one 
fermion per site:
\begin{equation}\label{eq:psivar}
|\Psi_v\rangle = {\cal P}_G |\Psi_{MF}\rangle,
\end{equation}
where ${\cal P}_G = \prod_i (n_{i,\uparrow}-n_{i,\downarrow})^2$ is the 
Gutzwiller projector, $n_{i,\sigma}=c^\dag_{i,\sigma} c_{i,\sigma}$ being the 
local density. Depending on the symmetry of the mean-field ansatz (e.g., the 
pattern of the $t_{i,j}$'s and the $\eta_{i,j}$'s), the projected state may 
describe {\it different} spin liquids, having for example $U(1)$ or $Z_2$ 
gauge structure and gapped or gapless spinon spectrum.~\cite{wen2002}

Here, we will consider a state that is obtained by taking a real pairing 
$\eta_{xy}$ (with $d_{xy}$ symmetry) on top of the $U(1)$ state with 
nearest-neighbor hopping $t$ and real pairing $\eta_{x^2-y^2}$ (with 
$d_{x^2-y^2}$ symmetry). The $d_{xy}$ term is crucial to break the $U(1)$ 
gauge symmetry down to $Z_2$. Restricting this coupling along the 
$(\pm 2,\pm 2)$ bonds implies {\it commensurate} Dirac points at 
$k=(\pm \pi/2, \pm \pi/2)$ in the mean-field spectrum; with this choice, 
the optimal variational state is found by projecting the ground state of 
${\cal H}_{MF}$.~\cite{nota1}

\begin{figure}
\includegraphics[width=\columnwidth]{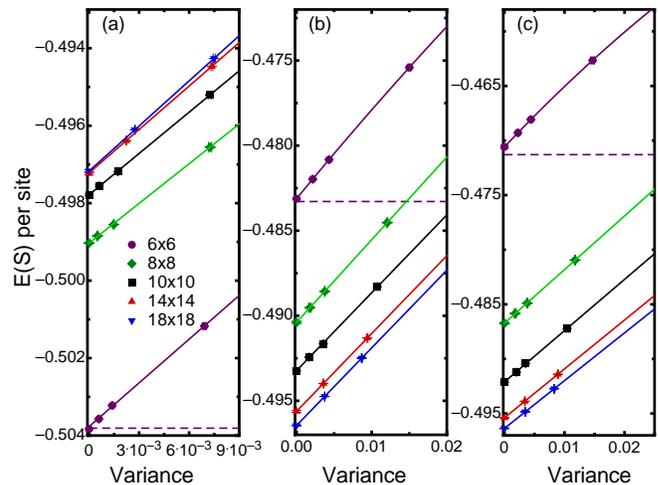}
\caption{\label{fig:accuracy}
(Color on-line) Energies per site for the $S=0$ ground state (a), the $S=1$ 
state with $k=(\pi,0)$ (b), and the $S=2$ with $k=(0,0)$ (c) versus the 
variance for $J_2/J_1=0.5$. The results with $p=0$, $1$, and $2$ are reported 
for $L=6$, $8$, and $10$, and with only $p=0$ and $p=1$ for $L=14$ and $L=18$. 
The variance extrapolated results are also shown.}
\end{figure}

\begin{figure}
\includegraphics[width=\columnwidth]{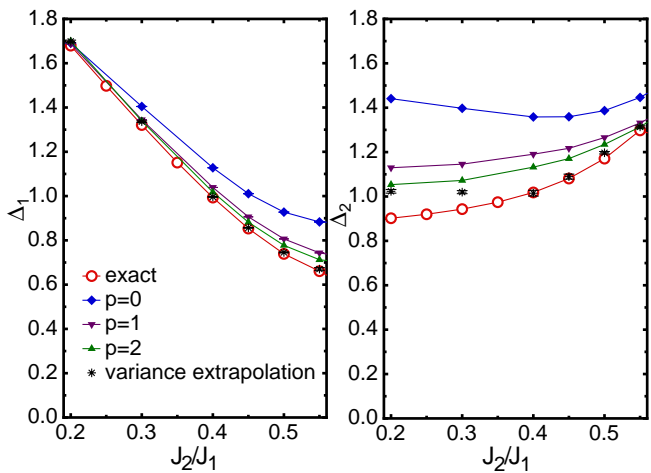}
\caption{\label{fig:6x6}
(Color on-line) Spin gap for the $S=1$ excitation at $k=(\pi,0)$ (left panel)
and the $S=2$ excitation with $k=(0,0)$ (right panel) for the $6 \times 6$
cluster. Results for $p=0$, $1$, and $2$ Lanczos steps are reported, together
with the extrapolated and the exact ones.}
\end{figure}

Besides the ansatz for the ground state, within this formalism, it is 
straightforward to have simple representations also for excited states. 
In this respect, it is useful to consider a particle-hole transformation for 
the down electrons on the mean-field Hamiltonian~(\ref{eq:meanfield}), i.e., 
$c^\dag_{i,\downarrow} \to c_{i,\downarrow}$, such that the transformed 
Hamiltonian conserves the total number of particles. Then, the ground state is 
obtained by filling the lowest $N$ orbitals, with suitable boundary conditions 
(either periodic or anti-periodic) in order to have a unique mean-field ground 
state. Spin excitations can be obtained by creating the appropriate Bogoliubov 
quasi-particles (spinons) and possibly switching boundary condition. By limiting
to states that can be constructed from a single determinant, here we consider a 
$S=2$ state with momentum $k=(0,0)$ and a $S=1$ state with $k=(\pi,0)$ or 
$(0,\pi)$. By performing Monte Carlo calculations, we are able to compute 
separately the energies of these three states, so to assess the spin gap of 
the $J_1{-}J_2$ model.

In order to systematically improve the variational wave functions, we can 
apply a number $p$ of Lanczos steps:
\begin{equation}
|\Psi_p\rangle = \left ( 1 + \sum_{m=1}^{p} \alpha_m {\cal H}^m \right ) 
|\Psi_v\rangle,
\end{equation}
where $\alpha_m$ are $p$ additional variational parameters. Clearly,
whenever $|\Psi_v\rangle$ is not orthogonal to the exact ground state,
$|\Psi_p\rangle$ converges to it for large $p$. Unfortunately, on large sizes,
only few steps can be efficiently afforded: here, we consider the case with
$p=1$ and $p=2$ ($p=0$ corresponds to the original variational wave 
function).~\cite{nota2}
Furthermore, an estimate of the exact energy may be obtained by the variance 
extrapolation. Indeed, for a systematically convergent sequence of states 
$|\Psi_p\rangle$ with energy $E_p$ and variance $\sigma_p^2$, it is easy to 
prove that $E_p \approx E_{\rm ex}+{\rm const} \times \sigma_p^2$, where 
$E_p=\langle \Psi_p|{\cal H}|\Psi_p\rangle/N$ and
$\sigma_p^2=(\langle \Psi_p|{\cal H}^2|\Psi_p\rangle-
\langle \Psi_p|{\cal H}|\Psi_p\rangle^2)/N$ are the energy and variance per 
site, respectively. Therefore, the exact energy $E_{\rm ex}$ may be extracted 
by fitting $E_p$ vs $\sigma_p^2$, for $p=0,1$, and $2$.

{\it Results --}
A systematic analysis shows that the best possible ansatz for the variational 
wave function of the form~(\ref{eq:psivar}) has a non-vanishing $d_{xy}$ pairing
in the whole regime $0.45 \lesssim J_2/J_1 \lesssim 0.6$. Here, both the static
structure factor $S(q)$ and the dimer-dimer correlations do not show any 
evidence for the occurrence of ordered states,~\cite{capriotti2001,book} in 
agreement with the DMRG results of Ref.~\onlinecite{jiang2012}. 

Let us start by showing the accuracy of our method for the ground state and 
the two excitations: $S=1$ at $k=(\pi,0)$ and $S=2$ at $k=(0,0)$. 
In Fig.~\ref{fig:accuracy}, we report calculations for $J_2/J_1=0.5$ and 
different sizes of the cluster. For $L=6$, where the exact results can be 
obtained by Lanczos diagonalizations, our extrapolations are extremely 
accurate. Moreover, for the ground state, our best variational $p=2$ 
state gives $E/J_1=-0.503571(3)$, while $E_{\rm ex}/J_1=-0.50381$; remarkably, 
the Lanczos step procedure remains effective even for larger sizes, the 
difference between the best variational state with $p=2$ and the extrapolated 
being very weakly size dependent (for $L=10$, the $p=2$ energy is 
$E/J_1=-0.497549(2)$, while the extrapolated one is $E/J_1=-0.49781(2)$). 
The same applies also for excited states, see Fig.~\ref{fig:accuracy}.
The almost perfect alignment of the Lanczos steps, together with the impressive
accuracy obtained up to relatively large clusters, clearly indicates that the
exact ground state should be essentially described by the starting $Z_2$ 
gapless state. 

\begin{figure}
\includegraphics[width=\columnwidth]{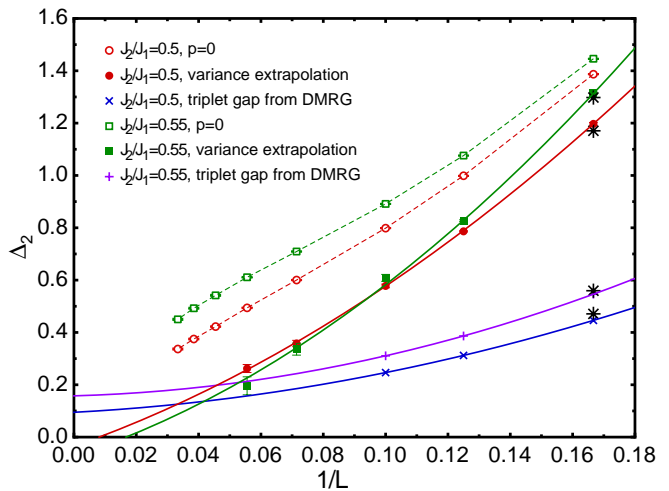}
\caption{\label{fig:gap2}
(Color on-line) The $S=2$ spin gap as a function of the system size for the
variational wave function and the Lanczos extrapolation for two values of
$J_2$. The thermodynamic extrapolations are consistent with a vanishing 
gap within the error bars, i.e., $\Delta_2=-0.04(5)$ and $-0.07(7)$ for 
$J_2/J_1=0.5$ and $0.55$, respectively. The DMRG results on $2L \times L$
cylinders (with open boundary conditions along $x$ and periodic along $y$)
for the $S=1$ excitation are also shown.~\cite{jiang2012} Exact results 
(stars) of the $S=2$ gap and the lowest $S=1$ gap on the $6 \times 6$ cluster 
(with periodic boundary conditions) are reported.}
\end{figure}

\begin{figure}
\includegraphics[width=\columnwidth]{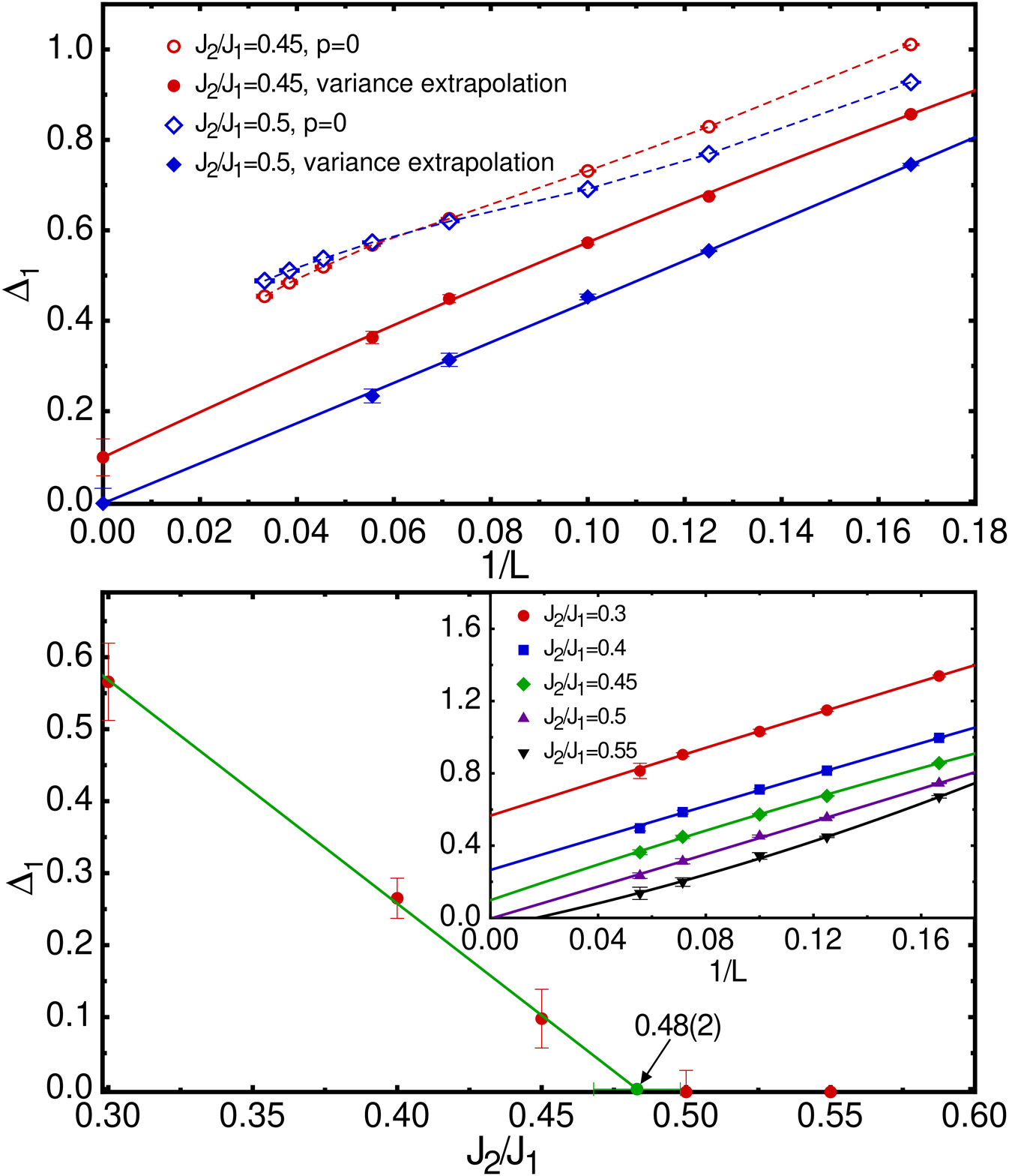}
\caption{\label{fig:gap1}
(Color on-line) The $S=1$ with $k=(\pi,0)$ spin gap as a function of the system
size for the variational wave function and the Lanczos extrapolation for two 
values of $J_2$ (upper panel). The behavior of the extrapolated gap as a 
function of $J_2/J_1$ is reported in the lower panel, the line is a guide to 
the eye. The Lanczos extrapolated gap as a function of $L$ for different 
values of $J_2$ are also reported in the inset.}
\end{figure}

In Fig.~\ref{fig:6x6}, we show the results for the $S=1$ spin gap $\Delta_1$ 
at $k=(\pi,0)$ and the $S=2$ spin gap $\Delta_2$ for the $6 \times 6$ cluster, 
in comparison with the exact results. Remarkably, our approach based upon a 
spin-liquid wave function gives excellent accuracy on $\Delta_1$ in the whole 
region $0.2 \le J_2/J_1 \le 0.55$. A similar accuracy is also obtained for
$\Delta_2$ in the strongly frustrated region (i.e., $0.4 \le J_2/J_1 \le 0.55$)
even though this is not a simple excitation since it involves four spinons. 
Instead, at $J_2/J_1 \simeq 0.6$ the accuracy deteriorates because a first-order
transition to the collinear magnetic state takes place in the thermodynamic 
limit:~\cite{doucot1988,jiang2012,schulz1996} in this region, a quasi-degeneracy
of levels in the energy spectrum occurs, leading to a reduced overlap between 
the variational wave function and the lowest exact 
eigenstate.~\cite{capriotti2001,book} 

Then, we consider larger cluster and perform a size scaling of the gaps, see
Figs.~\ref{fig:gap2} and~\ref{fig:gap1} for the $S=2$ and $S=1$ with 
$k=(\pi,0)$, respectively. For $L \ge 6$, the extrapolations obtained with 
two ($p=0$ and $1$) or three ($p=0$, $1$, and $2$) points are perfectly 
consistent (i.e., the three points lie along a straight line, see
Fig.~\ref{fig:accuracy}) Therefore, we perform the computationally demanding 
second Lanczos step only for relatively small clusters (up to the $L=10$), 
while we limit to the first Lanczos step for large clusters (up to the $L=18$). 

The $S=2$ gap is reported for two values of the frustrating ratio $J_2/J_1$, 
together with the $S=1$ gap obtained by DMRG calculations of 
Ref.~\onlinecite{jiang2012}. We find that the Lanczos step procedure clearly 
reduces the gap on any size. In contrast with the DMRG picture, we have a clear 
evidence that the spin gap closes when $L \to \infty$ for $J_2/J_1=0.5$ and 
$0.55$. Indeed, the values that we obtain in the thermodynamic limit are both 
compatible with a vanishing gap, i.e., $\Delta_2=-0.04(5)$ and $-0.07(7)$, 
see Fig.~\ref{fig:gap2}. We want to stress that our calculations are done on 
square clusters, having all the symmetries of the infinite lattice, and 
periodic boundary conditions, while DMRG calculations employed cylinders with 
$2L \times L$ sites and open boundary conditions along $x$. A possible
explanation for having a finite gap within DMRG is that this method favors
low-entangled states with finite gaps. On the contrary, our variational
approach is more flexible, allowing both gapped and gapless states. At the
pure $p=0$ variational level, the best wave function of the 
form~(\ref{eq:psivar}) is found to be gapless, its energy being the lowest one
among all states constructed from Schwinger bosons and Abrikosov fermions for
$0.45 \lesssim J_2/J_1 \lesssim 0.6$;~\cite{li2012} moreover, by  applying few 
Lanczos steps, the finite-size gap lowers with no evidence for a finite value
in the thermodynamic limit.

Finally, the $S=1$ gap with $k=(\pi,0)$ has been computed for various values 
of $J_2$ and cluster sizes, see Fig.~\ref{fig:gap1}. This gap is finite in the
N\'eel phase for small $J_2/J_1$, where the only gapless $S=1$ excitations 
have $k=(0,0)$ and $k=(\pi,\pi)$. Indeed, this is what is found for 
$J_2/J_1 \lesssim 0.48$ when the Lanczos extrapolation is considered, even if
the starting variational wave function is gapless before Gutzwiller projection.
Remarkably, in agreement with the theoretical picture of the Z2Azz13 spin 
liquid, this gap vanishes for the two cases we investigated within the 
spin liquid region: $J_2/J_1=0.5$ and $0.55$ ({\it before} the transition to 
the collinear magnetic phase, which occurs for $J_2/J_1 \gtrsim 0.6$). 
We expect that the $S=1$ gap at $k=(\pi,0)$ closes for $J_2 \to J_2^c$ with 
a non-trivial exponent smaller than one, which is however not possible to 
estimate with our numerical results. Nevertheless, by performing a linear fit 
of our data, we can obtain an upper bound of the N\'eel to spin liquid 
transition, which can be located at $J_2^c=0.48(2)$.

{\it Conclusions --}
In conclusion, by using a particularly accurate variational state and a 
procedure based upon the application of few Lanczos steps, we showed that it 
is possible to extract important information on the spin gap of frustrated 
spin models. In particular, we provided a solid evidence that the spin-liquid
phase of the $J_1{-}J_2$ model on the square lattice is gapless and may be 
very well described by using a Abrikosov-fermion mean field with a $Z_2$ gauge
structure and gapless spinons with four Dirac points at 
$k=(\pm \pi/2, \pm \pi/2)$. The latter statement is further supported by the 
occurrence of a vanishing $S=1$ gap at the non-trivial momenta $k=(\pi,0)$ 
and $(0,\pi)$. Our calculations give the first direct evidence for the existence
and the stability of highly-entangled gapless spin liquids in SU(2) spin models.

We thank Y. Iqbal and S.L. Sondhi for useful discussions. F.B. thanks 
P.A. Lee and H.-C. Jiang for very interesting discussions during the KITP 
program ``Frustrated Magnetism and Quantum Spin Liquids: From Theory and Models
to Experiments'' and partial support from the National Science Foundation under 
the Grant No. NSF PHY11-25915. We acknowledge support from PRIN 2010-11.

\newpage
\begin{table*}[!hbp]
\centering
\caption{p=0}
\begin{tabular}{|c|c|c|c|c|}
\hline
$\                      $ & $J_2/J_1=0.4 $ & $J_2/J_1=0.45$ & $J_2/J_1=0.5 $ & $J_2/J_1=0.55$   \\ \hline
$L=6\ \ \            S=0$ & $-0.52715(1) $ & $-0.51364(1) $ & $-0.50117(1) $ & $-0.48992(1) $ \\
$\ \ \ \ \ \ \ \ \ \ S=1$ & $-0.49582(1) $ & $-0.48557(1) $ & $-0.47541(1) $ & $-0.46538(2) $ \\
$\ \ \ \ \ \ \ \ \ \ S=2$ & $\           $ & $\           $ & $-0.46265(2) $ & $-0.44974(2) $ \\ \hline
$L=8\ \ \            S=0$ & $-0.52302(1) $ & $-0.50930(1) $ & $-0.49656(1) $ & $-0.48487(1) $ \\
$\ \ \ \ \ \ \ \ \ \ S=1$ & $-0.50835(1) $ & $-0.49635(1) $ & $-0.48453(1) $ & $-0.47299(1) $ \\
$\ \ \ \ \ \ \ \ \ \ S=2$ & $\           $ & $\           $ & $-0.48095(1) $ & $-0.46806(1) $ \\ \hline
$L=10\               S=0$ & $-0.52188(1) $ & $-0.50811(1) $ & $-0.49521(1) $ & $-0.48335(1) $ \\
$\ \ \ \ \ \ \ \ \ \ S=1$ & $-0.51362(1) $ & $-0.50080(1) $ & $-0.48830(1) $ & $-0.47625(1) $ \\
$\ \ \ \ \ \ \ \ \ \ S=2$ & $\           $ & $\           $ & $-0.48722(1) $ & $-0.47443(1) $ \\ \hline
$L=14\               S=0$ & $-0.52124(1) $ & $-0.50745(1) $ & $-0.49447(1) $ & $-0.48242(1) $ \\
$\ \ \ \ \ \ \ \ \ \ S=1$ & $-0.51772(1) $ & $-0.50425(1) $ & $-0.49131(1) $ & $-0.47904(1) $ \\
$\ \ \ \ \ \ \ \ \ \ S=2$ & $\           $ & $\           $ & $-0.49141(1) $ & $-0.47880(1) $ \\ \hline
$L=18\               S=0$ & $-0.52107(1) $ & $-0.50728(1) $ & $-0.49426(1) $ & $-0.48215(1) $ \\
$\ \ \ \ \ \ \ \ \ \ S=1$ & $-0.51921(1) $ & $-0.50553(1) $ & $-0.49249(1) $ & $-0.48018(1) $ \\
$\ \ \ \ \ \ \ \ \ \ S=2$ & $\           $ & $\           $ & $-0.49274(1) $ & $-0.48026(1) $ \\ \hline
\end{tabular}
\end{table*}

\begin{table*}[!hbp]
\centering
\caption{p=1}
\begin{tabular}{|c|c|c|c|c|}
\hline
$\                      $ & $J_2/J_1=0.4 $ & $J_2/J_1=0.45$ & $J_2/J_1=0.5 $ & $J_2/J_1=0.55$ \\ \hline
$L=6\ \ \            S=0$ & $-0.52928(1) $ & $-0.51538(1) $ & $-0.50323(1) $ & $-0.49303(1) $ \\
$\ \ \ \ \ \ \ \ \ \ S=1$ & $-0.50042(1) $ & $-0.49020(1) $ & $-0.48082(1) $ & $-0.47238(1) $ \\
$\ \ \ \ \ \ \ \ \ \ S=2$ & $\           $ & $\           $ & $-0.46807(1) $ & $-0.45605(1) $ \\ \hline
$L=8\ \ \            S=0$ & $-0.52501(1) $ & $-0.51101(1) $ & $-0.49855(1) $ & $-0.48777(1) $ \\
$\ \ \ \ \ \ \ \ \ \ S=1$ & $-0.51157(1) $ & $-0.49963(1) $ & $-0.48857(1) $ & $-0.47847(1) $ \\
$\ \ \ \ \ \ \ \ \ \ S=2$ & $\           $ & $\           $ & $-0.48489(1) $ & $-0.47305(1) $ \\ \hline
$L=10\               S=0$ & $-0.52368(1) $ & $-0.50973(1) $ & $-0.49718(1) $ & $-0.48622(1) $ \\
$\ \ \ \ \ \ \ \ \ \ S=1$ & $-0.51610(1) $ & $-0.50344(1) $ & $-0.49165(1) $ & $-0.48090(1) $ \\
$\ \ \ \ \ \ \ \ \ \ S=2$ & $\           $ & $\           $ & $-0.49041(1) $ & $-0.47867(1) $ \\ \hline
$L=14\               S=0$ & $-0.52287(1) $ & $-0.50899(1) $ & $-0.49638(1) $ & $-0.48519(1) $ \\
$\ \ \ \ \ \ \ \ \ \ S=1$ & $-0.51966(1) $ & $-0.50632(1) $ & $-0.49398(1) $ & $-0.48270(1) $ \\
$\ \ \ \ \ \ \ \ \ \ S=2$ & $\           $ & $\           $ & $-0.49387(1) $ & $-0.48221(1) $ \\ \hline
$L=18\               S=0$ & $-0.52259(1) $ & $-0.50874(1) $ & $-0.49611(1) $ & $-0.48475(1) $ \\
$\ \ \ \ \ \ \ \ \ \ S=1$ & $-0.52083(5) $ & $-0.50137(1) $ & $-0.49475(1) $ & $-0.48327(1) $ \\
$\ \ \ \ \ \ \ \ \ \ S=2$ & $\           $ & $\           $ & $-0.49485(1) $ & $-0.48319(1) $ \\ \hline
\end{tabular}
\end{table*}

\begin{table*}[!hbp]
\centering
\caption{p=2}
\begin{tabular}{|c|c|c|c|c|}
\hline
$\                      $ & $J_2/J_1=0.4 $ & $J_2/J_1=0.45$ & $J_2/J_1=0.5 $ & $J_2/J_1=0.55$   \\ \hline
$L=6\ \ \            S=0$ & $-0.52957(1) $ & $-0.51558(1) $ & $-0.50357(1) $ & $-0.49399(1) $ \\
$\ \ \ \ \ \ \ \ \ \ S=1$ & $-0.50130(1) $ & $-0.49108(1) $ & $-0.48197(1) $ & $-0.47419(1) $ \\
$\ \ \ \ \ \ \ \ \ \ S=2$ & $\           $ & $\           $ & $-0.46929(1) $ & $-0.45750(1) $ \\ \hline
$L=8\ \ \            S=0$ & $-0.52539(1) $ & $-0.51125(1) $ & $-0.49886(1) $ & $-0.48841(2) $ \\
$\ \ \ \ \ \ \ \ \ \ S=1$ & $-0.51224(2) $ & $-0.50033(1) $ & $-0.48952(1) $ & $-0.48008(4) $ \\
$\ \ \ \ \ \ \ \ \ \ S=2$ & $\           $ & $\           $ & $-0.48583(4) $ & $-0.47443(2) $ \\ \hline
$L=10\               S=0$ & $-0.5240(1)  $ & $-0.51001(1) $ & $-0.49755(1) $ & $-0.48693(3) $ \\
$\ \ \ \ \ \ \ \ \ \ S=1$ & $-0.51671(7) $ & $-0.50398(1) $ & $-0.49243(1) $ & $-0.4825(2)  $ \\
$\ \ \ \ \ \ \ \ \ \ S=2$ & $\           $ & $\           $ & $-0.49121(3) $ & $-0.4800(2)  $ \\ \hline
$L=14\               S=0$ & $\ $ & $\ $ & $\ $ & $\ $ \\
$\ \ \ \ \ \ \ \ \ \ S=1$ & $\ $ & $\ $ & $\ $ & $\ $ \\
$\ \ \ \ \ \ \ \ \ \ S=2$ & $\ $ & $\ $ & $\ $ & $\ $ \\ \hline
$L=18\               S=0$ & $\ $ & $\ $ & $\ $ & $\ $ \\
$\ \ \ \ \ \ \ \ \ \ S=1$ & $\ $ & $\ $ & $\ $ & $\ $ \\
$\ \ \ \ \ \ \ \ \ \ S=2$ & $\ $ & $\ $ & $\ $ & $\ $ \\ \hline
\end{tabular}
\end{table*}

\begin{table*}[!hbp]
\centering
\caption{extrapolation}
\begin{tabular}{|c|c|c|c|c|}
\hline
$\                      $ & $J_2/J_1=0.4 $ & $J_2/J_1=0.45$ & $J_2/J_1=0.5 $ & $J_2/J_1=0.55$   \\ \hline
$L=6\ \ \            S=0$ & $-0.52972(1) $ & $-0.51566(1) $ & $-0.50382(1) $ & $-0.49521(7) $ \\
$\ \ \ \ \ \ \ \ \ \ S=1$ & $-0.50204(5) $ & $-0.49187(4) $ & $-0.48312(6) $ & $-0.4766(1)  $ \\
$\ \ \ \ \ \ \ \ \ \ S=2$ & $\           $ & $\           $ & $-0.4706(1)  $ & $-0.4587(1)  $ \\ \hline
$L=8\ \ \            S=0$ & $-0.52556(1) $ & $-0.51140(1) $ & $-0.49906(1) $ & $-0.48894(3) $ \\
$\ \ \ \ \ \ \ \ \ \ S=1$ & $-0.51282(1) $ & $-0.50085(1) $ & $-0.49039(2) $ & $-0.48194(3) $ \\
$\ \ \ \ \ \ \ \ \ \ S=2$ & $\           $ & $\           $ & $-0.48677(1) $ & $-0.47602(3) $ \\ \hline
$L=10\               S=0$ & $-0.52429(2) $ & $-0.51017(2) $ & $-0.49781(2) $ & $-0.48766(6) $ \\
$\ \ \ \ \ \ \ \ \ \ S=1$ & $-0.51718(3) $ & $-0.50445(3) $ & $-0.49329(5) $ & $-0.4842(1)  $ \\
$\ \ \ \ \ \ \ \ \ \ S=2$ & $\           $ & $\           $ & $-0.49203(5) $ & $-0.48157(8) $ \\ \hline
$L=14\               S=0$ & $-0.52351(2) $ & $-0.50953(1) $ & $-0.49722(2) $ & $-0.48696(5) $ \\
$\ \ \ \ \ \ \ \ \ \ S=1$ & $-0.52052(2) $ & $-0.50724(3) $ & $-0.49562(5) $ & $-0.48594(7) $ \\
$\ \ \ \ \ \ \ \ \ \ S=2$ & $\           $ & $\           $ & $-0.49539(4) $ & $-0.48524(9) $ \\ \hline
$L=18\               S=0$ & $-0.52333(1) $ & $-0.50940(1) $ & $-0.49717(2) $ & $-0.48698(5) $ \\
$\ \ \ \ \ \ \ \ \ \ S=1$ & $-0.52180(4) $ & $-0.50828(3) $ & $-0.49645(3) $ & $-0.48656(5) $ \\
$\ \ \ \ \ \ \ \ \ \ S=2$ & $\           $ & $\           $ & $-0.49636(3) $ & $-0.48638(5) $ \\ \hline
\end{tabular}
\end{table*}


\begin{thebibliography}{99}
\bibitem{balents2010} L. Balents, Nature {\bf 464}, 199 (2010).
\bibitem{moessner2001} R. Moessner and S. Sondhi, \prl {\bf 86}, 1881 (2001).
\bibitem{balents2002} L. Balents, M.P.A. Fisher, and S.M. Girvin, \prb {\bf 65},
   224412 (2002).
\bibitem{kitaev2003} A.Y. Kitaev, Ann. Phys. {\bf 303}, 2 (2003).
\bibitem{kitaev2005} A.Y. Kitaev, Ann. Phys. {\bf 321}, 2 (2005).
\bibitem{wen2002} X.-G. Wen, \prb {\bf 65}, 165113 (2002).
\bibitem{chen2011a} X. Chen, Z.-C. Gu, and X.-G. Wen, \prb {\bf 83}, 035107 
   (2011); N. Schuch, D. Perez-García, and I. Cirac, \prb {\bf 84}, 165139 
   (2011).
\bibitem{chen2011b} X. Chen, Z.-C. Gu, Z.-X. Liu, and X.-G. Wen, \prb {\bf 87},
   155114 (2013).
\bibitem{hermele2012} A.M. Essin and M. Hermele, \prb {\bf 87}, 104406 (2013).
\bibitem{ran2012} A. Mesaros and Y. Ran, \prb {\bf 87}, 155115 (2013).
\bibitem{capriotti2001} L. Capriotti, F. Becca, A. Parola, and S. Sorella, 
   \prl {\bf 87}, 097201 (2001).
\bibitem{yan2011} S. Yan, D. Huse, and S. White, Science {\bf 332}, 1173 (2011).
\bibitem{jiang2012} H.-C. Jiang, H. Yao, and L. Balents, \prb {\bf 86}, 024424
   (2012).
\bibitem{iqbal2013} Y. Iqbal, F. Becca, S. Sorella, and D. Poilblanc, \prb 
   {\bf 87}, 060405 (2013).
\bibitem{meng2010} Z.Y. Meng, T.C. Lang, S. Wessel, F.F. Assaad, and 
   A. Muramatsu, Nature {\bf 464}, 847 (2010).
\bibitem{sorella2012} S. Sorella, Y. Otsuka, and S. Yunoki, Sci. Rep. {\bf 2}, 
   992 (2012).
\bibitem{sandvik1997} A.W. Sandvik, \prb {\bf 56}, 11678 (1997).
\bibitem{calandra1998} M. Calandra Buonaura and S. Sorella, \prb {\bf 57}, 11446
   (1998).
\bibitem{doucot1988} P. Chandra and B. Doucot, \prb {\bf 38}, 9335 (1988).
\bibitem{read1989} N. Read and S. Sachdev, \prl {\bf 62}, 1694 (1989).
\bibitem{gelfand1989} M.P. Gelfand, R.R.P. Singh, and D.A. Huse, \prb {\bf 40},
   10801 (1989).
\bibitem{figueirido1990} F. Figueirido, A. Karlhede, S. Kivelson, S. Sondhi, 
   M. Rocek, and D.S. Rokhsar, \prb {\bf 41}, 4619 (1990).
\bibitem{darradi2008} R. Darradi, O. Derzhko, R. Zinke, J. Schulenburg, 
   S E. Kruger, and J. Richter, \prb {\bf 78}, 214415 (2008).
\bibitem{mambrini2006} M. Mambrini, A. Lauchli, D. Poilblanc, and F. Mila, \prb
   {\bf 74}, 144422 (2006).
\bibitem{arovas1988} D.P. Arovas and A. Auerbach, \prb {\bf 38}, 316 (1988).
\bibitem{li2012} T. Li, F. Becca, W.-J. Hu, and S. Sorella, \prb {\bf 86},
   075111 (2012).
\bibitem{wang2013} L. Wang, D. Poilblanc, Z.-C. Gu, X.-G. Wen, and 
   F. Verstraete, arXiv:1301.4492.
\bibitem{baskaran1987} G. Baskaran, Z. Zou, and P.W. Anderson, Solid State 
   Commun. {\bf 63}, 973 (1987); G. Baskaran and P.W. Anderson, \prb {\bf 37},
   580 (1988).
\bibitem{nota1} In this approach, all pairing terms are variational
   parameters that can be optimized in order to lower the total energy by 
   using the method described by S. Sorella, \prb {\bf 71}, 241103 (2005).
   For $L>6$, we considered $\eta_{x^2-y^2}$ terms for bonds $(1,0)$, $(2,1)$,
   and $(3,0)$ (and symmetry related ones) and $\eta_{xy}$ terms for 
   $(\pm 2,\pm 2)$ bonds. 
\bibitem{nota2} Within quantum Monte Carlo calculations, in order to have a 
   stable and accurate algorithm for $p$ Lanczos steps, it is important to 
   adopt a suitable regularization scheme to avoid vanishingly small 
   determinants. Here, we considered to sample configurations $|x\rangle$ such 
   that: $\langle x|{\cal H}|\Psi_v\rangle/\langle x|\Psi_v\rangle>N/\epsilon$,
   with $\epsilon=10^{-6}$ (no appreciable change is found with $\epsilon$ 
   ranging from $10^{-8}$ to $10^{-4}$); W.-J. Hu, F. Becca, A. Parola, and 
   S. Sorella, in preparation.
\bibitem{book} F. Becca, L. Capriotti, A. Parola, and S. Sorella, 
   Springer Ser. Solid-State Sci. {\bf 164}, 379 (2011).
\bibitem{schulz1996} H.J. Schulz, T. Ziman, and D. Poilblanc, J. Phys. I 
   {\bf 6}, 675 (1996).
\end{thebibliography}
\end{document}